\documentclass[12pt]{article}
\usepackage{amsmath}
\usepackage{bm}
\usepackage{amssymb}
\usepackage{amsthm}
\usepackage{longtable}
\usepackage{amsfonts}
\usepackage{geometry}
\usepackage{multirow}
\usepackage{multicol}
\usepackage{graphicx}
\usepackage{epsfig}
\usepackage{epstopdf}
\usepackage{subfigure}
\usepackage{color}
\usepackage{ulem}
\usepackage{booktabs}
\usepackage{rotating}
\usepackage[toc,page]{appendix}
\usepackage{cite}
\usepackage{cases}
\usepackage{makecell}
\renewcommand{\theequation}{\arabic{section}.\arabic{equation}}

\begin{document}



\def\a{\alpha}
\def\b{\beta}
\def\d{\delta}
\def\e{\epsilon}
\def\g{\gamma}
\def\h{\mathfrak{h}}
\def\k{\kappa}
\def\l{\lambda}
\def\o{\omega}
\def\p{\wp}
\def\r{\rho}
\def\t{\tau}
\def\s{\sigma}
\def\z{\zeta}
\def\x{\xi}
\def\V={{{\bf\rm{V}}}}
 \def\A{{\cal{A}}}
 \def\B{{\cal{B}}}
 \def\C{{\cal{C}}}
 \def\D{{\cal{D}}}
\def\G{\Gamma}
\def\K{{\cal{K}}}
\def\O{\Omega}
\def\R{\bar{R}}
\def\T{{\cal{T}}}
\def\L{\Lambda}
\def\f{E_{\tau,\eta}(sl_2)}
\def\E{E_{\tau,\eta}(sl_n)}
\def\Zb{\mathbb{Z}}
\def\Cb{\mathbb{C}}

\def\R{\overline{R}}

\def\beq{\begin{equation}}
\def\eeq{\end{equation}}
\def\bea{\begin{eqnarray}}
\def\eea{\end{eqnarray}}
\def\ba{\begin{array}}
\def\ea{\end{array}}
\def\no{\nonumber}
\def\le{\langle}
\def\re{\rangle}
\def\lt{\left}
\def\rt{\right}

\baselineskip=20pt

\newfont{\elevenmib}{cmmib10 scaled\magstep1}
\newcommand{\preprint}{
   \begin{flushleft}
   \end{flushleft}\vspace{-1.3cm}
   \begin{flushright}\normalsize
   \end{flushright}}
\newcommand{\Title}[1]{{\baselineskip=26pt
   \begin{center} \Large \bf #1 \\ \ \\ \end{center}}}

\newcommand{\Author}{\begin{center}
	\large \bf
	Pengcheng Lu${\,}^{a,}$\footnote{pengcheng.lu@uq.edu.au},
    Junpeng Cao${\,}^{b,e,f,g}$,
    Wen-Li Yang${\,}^{c,g,h}$,
    Ian Marquette${\,}^{d}$
    and Yao-Zhong Zhang${\,}^{a}$
\end{center}}

\newcommand{\Address}{\begin{center}

    ${}^a$ School of Mathematics and Physics, The University of Queensland, Brisbane, QLD 4072, Australia\\
	${}^b$ Beijing National Laboratory for Condensed Matter Physics, Institute of Physics, Chinese Academy of Sciences, Beijing 100190, China\\
    ${}^c$ Institute of Modern Physics, Northwest University, Xian 710127, China\\
    ${}^d$ Department of Mathematical and Physical Sciences, La Trobe University, Bendigo, VIC 3552, Australia\\
    ${}^e$ School of Physical Sciences, University of Chinese Academy of Sciences, Beijing 100049, China\\
	${}^f$ Songshan Lake Materials Laboratory, Dongguan, Guangdong 523808, China\\
	${}^g$ Peng Huanwu Center for Fundamental Theory, Xian 710127, China\\
	${}^h$ Shaanxi Key Laboratory for Theoretical Physics Frontiers, Xian 710127, China
\end{center}}

\Title{$T$-$W$ relation and free energy of the antiperiodic $XXZ$ chain with $\eta=\frac{i\pi}{3}$ at a finite temperature} \Author

\Address
\vspace{1cm}

\begin{abstract}
We study the thermodynamics of the antiperiodic XXZ chain with anisotropy parameter $\eta=\frac{i\pi}{3}$ by means of the $t-W$ scheme. We parameterize the eigenvalues of both the transfer matrix and the corresponding fused transfer matrix by their zero points instead of Bethe roots. Based on the patterns of the zero points distribution and the reconstructed entropy, we obtain the nonlinear integral equations (NLIEs) describing the thermodynamics of the model and compute its free energy at a finite temperature.

\vspace{1truecm} \noindent {\it PACS:}
75.10.Pq, 03.65.Vf, 71.10.Pm

\end{abstract}
\newpage

\section{Introduction}
\label{intro} \setcounter{equation}{0}

The XXZ spin chain with twisted boundary condition is interesting and important because of its topological properties \cite{NPB1995461,JPA19952759,NPB2008524,JPA2009195008,JPA2011015001,NPB2013397} such as the confined spinon excitations\cite{PRL2013137201}, topological translation symmetry\cite{PRB2020085115}, and so on. This model is a typical quantum integrable
model without $U(1)$ symmetry and plays a significant role in the recent studies of non-equilibrium statistical physics \cite{Vanicat18,Frassek20,Chen20,Godreau20}, condensed matter physics \cite{Andrei2}, cold atom physics \cite{Bastianello18,Mestyan19} and AdS/CFT correspondence \cite{Fontanella17,Jiang20,Leeuw21}. Due to the lack of $U(1)$ symmetry,
the conventional coordinate or algebraic Bethe ansatz \cite{ZP1931205,CUP2014,SPD1978902,RMS197911,CUP1993} does not work for this case. In \cite{S2015,NPB201570,JSM2015P05014},  exact solutions of the system including energy spectrum and helical eigenstates were obtained by means of the off-diagonal Bethe ansatz \cite{PRL2013137201}. However, due to the fact that the Bethe roots satisfy inhomogeneous Bethe ansatz equations (BAEs), the traditional thermodynamic Bethe ansatz (TBA) \cite{PRL19711301,TP1971401,CUP1005} is not applicable and it is a challenging problem to obtain the thermodynamic limit of the model.

Recently, a novel $t-W$ scheme\cite{PRB2020085115,PRB2021L220401} has been proposed for calculating physical quantities of integrable models without $U(1)$ symmetry. The key point of this scheme is parameterizing the eigenvalues of the transfer matrix by its zero points. By substituting the zero points into the single $t-W$ relation constructed from the fusion of the transfer matrix, homogeneous zero points constraint equations can be obtained. This way one can define densities of the states and derive exact results in the thermodynamic limit.
Using the $t-W$ method, the exact ground state energy and elementary excitations for the antiperiodic XXZ spin chain were obtained.

In this paper, we study the thermodynamics of this model with anisotropy parameter $\eta=\frac{i\pi}{3}$ at a finite temperature. We obtain the $t-W$ relation and the homogeneous zero points constraint equations satisfied by the eigenvalues of the associated transfer matrices. For the $\eta=\frac{i\pi}{3}$ case, the patterns of the zero points distribution for a state are determined by solving the constraint equations. Based on these, we obtain the NLIEs describing the thermodynamics of the model and calculate its free energy at a finite temperature.

The paper is organized as follows.  Section 2 introduces the antiperiodic XXZ spin chain and shows its integrability. In Section 3, the $t-W$ relations for the transfer
matrix and the corresponding eigenvalues are constructed based on the fusion method. In section 4, we derive the exact solutions of the system. The eigenvalues of the transfer matrix and the corresponding fused transfer matrix are parameterized by their zero points. The patterns of the zero points with anisotropy parameter $\eta=\frac{i\pi}{3}$ for any states are obtained. In section 5, the NLIEs  and  free energy describing the thermodynamics of the antiperiodic XXZ spin chain with anisotropy parameter $\eta=\frac{i\pi}{3}$ are derived.  We summarize our results in Section 6. Some supporting materials are given in appendices A and B.


\section{Integrability of the model}
\label{XXZ} \setcounter{equation}{0}
The antiperiodic XXZ spin chain is characterized by the Hamiltonian \cite{PR1121958}
\begin{eqnarray}
H = \sum_{n=1}^N
  (\sigma_n^x\sigma_{n+1}^x+\sigma_n^y\sigma_{n+1}^y
  +\cosh\eta \,\sigma_n^z\sigma_{n+1}^z), \label{xxzh}
\end{eqnarray}
with the twisted boundary condition
\bea
\sigma^{\alpha}_{1+N}=\sigma^{x}_{1}\sigma^{\alpha}_{1}\sigma^{x}_{1},\quad {\rm  for}\quad \alpha=x,y,z,\label{BC}
\eea
where $N$ is the number of sites, and $\sigma^x,\,\sigma^y,\,\sigma^z$ are the usual Pauli matrices.

The integrability of the model can be established from the well-known six-vertex $R$-matrix
\bea
 R_{12}(u)=\frac{1}{\sinh \eta}\lt(\begin{array}{llll}\sinh(u+\eta)&&&\\&\sinh u&\sinh\eta&\\
&\sinh\eta&\sinh u&\\&&&\sinh(u+\eta)\end{array}\rt).
\label{r-matrix} \eea
Here the generic complex number $\eta$ is the
crossing parameter which provides the $z$-direction coupling constant of the Hamiltonian (\ref{xxzh}). The $R$-matrix satisfies the quantum Yang-Baxter
equation (QYBE) \cite{PRL191967, Baxter1982},
\begin{eqnarray}
\hspace{-0.2truecm}R_{12}(u_1-u_2)R_{13}(u_1-u_3)R_{23}(u_2-u_3)=R_{23}(u_2-u_3)R_{13}(u_1-u_3)R_{12}(u_1-u_2).\label{QYB}
\end{eqnarray}
The transfer matrix $t(u)$ of the antiperiodic XXZ spin chain is constructed by the $R$-matrix as
 \begin{eqnarray}
 t(u)=tr_0\lt\{\sigma^{x}_{0}R_{0N}(u)\cdots R_{01}(u)\rt\},\label{trans-per}
 \end{eqnarray}
where $tr_0$ denotes trace over the
``auxiliary space" $0$. The expression (\ref{r-matrix}) of the $R$-matrix $R(u)$ and the transfer matrix (\ref{trans-per}) imply that
\bea\label{expansion}
t(u)=t_1e^{(N-1)u}+t_2e^{(N-3)u}+\cdots+t_Ne^{-(N-1)u},
\eea
where $\{t_j|j=1,\cdots,N\}$ are the expansion coefficients of $t(u)$. The Hamiltonian (\ref{xxzh}) with twisted boundary condition (\ref{BC}) is given by
\begin{eqnarray}
H=2\sinh\eta \,\frac{\partial \ln t(u)}{\partial
u}\Big|_{u=0}-N\cosh\eta.\label{ham}
\end{eqnarray}
It has been demonstrated that the transfer matrices with different spectral
parameters satisfy the commutation relation $[t(u),t(v)]=0$  \cite{CUP1993}.
Thus the integrability of the antiperiodic XXZ spin chain (\ref{xxzh})-(\ref{BC}) can be established by the transfer matrix $t(u)$ (\ref{trans-per}).


\section{$T-W$ relation  }
\label{t-W-operator}
\setcounter{equation}{0}

Now, we construct the $t-W$ relation of the transfer matrix and the corresponding eigenvalues by using the fusion method.

Let us take the product of the transfer matrices $t(u)$ and $t(u-\eta)$
\bea
t(u)t(u-\eta)&=&tr_{12}\lt\{\sigma^{x}_{1}\sigma^{x}_{2}T_2(u)\,T_1(u-\eta)\rt\}\no\\[4pt]
&=&tr_{12}\lt\{\sigma^{x}_{1}\sigma^{x}_{2}T_2(u)\,T_1(u-\eta)(P^{(-)}_{12}+P^{(+)}_{12})\rt\}\no\\[4pt]
&=&tr_{12}\lt\{P^{(-)}_{12}\sigma^{x}_{1}\sigma^{x}_{2}T_2(u)\,T_1(u-\eta)P^{(-)}_{12}\rt\}\no\\[4pt]
&&\quad\quad+tr_{12}\lt\{P^{(+)}_{12}\sigma^{x}_{1}\sigma^{x}_{2}T_2(u)\,T_1(u-\eta)P^{(+)}_{12}\rt\},\label{Proof}
\eea
where $P_{12}^{(\pm)}$ are the (anti)symmetric projection operators with the forms of $P_{12}^{(\pm)}=\frac{1}{2}(1\pm P_{12})$. The operator $P_{12}$ is the  permutation operator between two spaces. Keeping in mind of the fact ${\rm rank}(P^{(-)})=1$ and with the help of the fusion of $R$-matrix \cite{Kul81,Kir86},
we can show that the transfer matrix $t(u)$ satisfies the fusion relation \cite{Res83,wly06}
\bea
t(u)\,t(u-\eta)=-a(u)\,d(u-\eta)\times {\rm id}+d(u)\,\mathbb{W}(u),\label{t-W-relation-op}
\eea
where $a(u)$ and $d(u)$ are defined as
\begin{eqnarray}
&&a(u)=
\frac{\sinh^N(u+\eta)} {\sinh^N\eta},\quad d(u)=a(u-\eta)=
\frac{\sinh^N u} {\sinh^N\eta},\label{a-d-functions}
\end{eqnarray}
and $\mathbb{W}(u)$ is a fused transfer matrix from $t(u)$ (\ref{trans-per}), which is an $N$-degree operator-valued trigonometric polynomial of $u$ (The proof of the above relation (\ref{t-W-relation-op}) is relegated to Appendix A.).  Moreover, the transfer matrices $t(u)$ and $\mathbb{W}(u)$ with different spectral parameters satisfy the following commutative relation
\bea
[t(u),\,t(v)]=[t(u),\,\mathbb{W}(v)]=[\mathbb{W}(u),\,\mathbb{W}(v)]=0,\label{Communtivity}
\eea
which implies that the transfer matrices $t(u)$ has common eigenstates with $\mathbb{W}(u)$. Let us set $|\Psi\rangle$ as a common eigenstate with the eigenvalues $\Lambda(u)$ and $W(u)$, i.e.,
\bea
t(u)\,|\Psi\rangle=\Lambda(u)\,|\Psi\rangle,\qquad \mathbb{W}(u)\,|\Psi\rangle=W(u)\,|\Psi\rangle.\no
\eea
Then from the construction of the transfer matrices, we have that
\bea
&&\mbox{$\Lambda(u)$, as a function of $u$, is a trigonometric  polynomial of degree $N-1$}, \label{Elliptic-Poly-1}\\
&&\mbox{$W(u)$, as a function of $u$, is a trigonometric polynomial of degree $N$}.\label{Elliptic-Poly-2}
\eea

Moreover, from the fusion relation (\ref{t-W-relation-op}) we can show that the corresponding eigenvalues $\Lambda(u)$ and $W(u)$ satisfy the $t-W$ relation
\bea
\Lambda(u)\,\Lambda(u-\eta)=-a(u)\,d(u-\eta)+d(u)\,W(u).\label{t-W-relation-Eigen}
\eea


\section{Exact solution to the antiperiodic XXZ spin chain}
\label{XXZclosedchain} \setcounter{equation}{0}

The expansion expressions (\ref{expansion}) and (\ref{t-W-relation-Eigen}) allow us to express the eigenvalues $\Lambda(u)$ and $W(u)$ in terms of their $N-1$ zero points $z$-roots $\{z_j|j=1,\cdots,N-1\}$ and $N$ zero points $w$-roots $\{w_l|l=1,\cdots,N\}$) respectively
\bea
\hspace{-1.42truecm}&&\Lambda(u)= \Lambda_{0}\,
\prod_{j=1}^{N-1}\frac{\sinh(u-z_j+\frac{\eta}{2})}{\sinh\eta}, \label{Expansion-3}\\[4pt]
\hspace{-1.42truecm}&&W(u)=W_{0}\,
\prod_{l=1}^N \frac{\sinh(u-w_l)}{\sinh\eta},\,\label{Expansion-4}
\eea
where the coefficient $\Lambda_{0}$ can be determined by putting $u=0$ in (\ref{t-W-relation-Eigen}) as
\bea\label{Lam0}
\Lambda^{2}_{0}\,\prod_{j=1}^{N-1}\sinh(z_j-\frac{\eta}{2})\sinh(z_j+\frac{\eta}{2})=(-1)^{N-1}.
\eea
Since $\Lambda(u)$ is a degree $N-1$ trigonometric polynomial of $u$, the leading terms in the right hand side of (\ref{t-W-relation-Eigen}) must be zero. Therefore, the coefficient $W_{0}$ and the $w$-roots $\{w_l|l=1,\cdots,N\}$ must satisfy the following constraint
\bea\label{W0}
W_0\,e^{\pm\sum_{l=1}^N w_l}=1,\quad {\rm or}\quad W_0^2=1\quad {\rm and} \quad \sum_{l=1}^N w_l=0\quad {\rm mod}(i\pi).
\eea
\begin{figure}[htbp]
  \centering
  \subfigure{\label{fig1:subfig:a} 
    \includegraphics[scale=0.55]{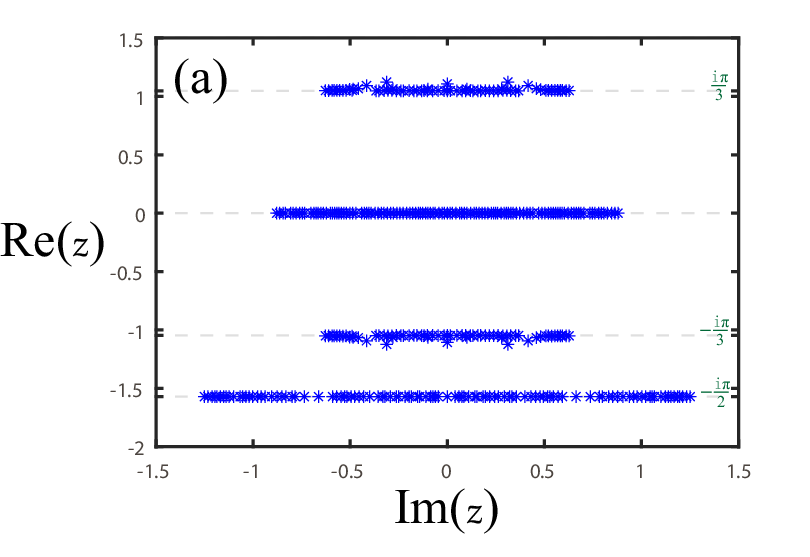}
    }
  \subfigure{\label{fig1:subfig:b} 
    \includegraphics[scale=0.55]{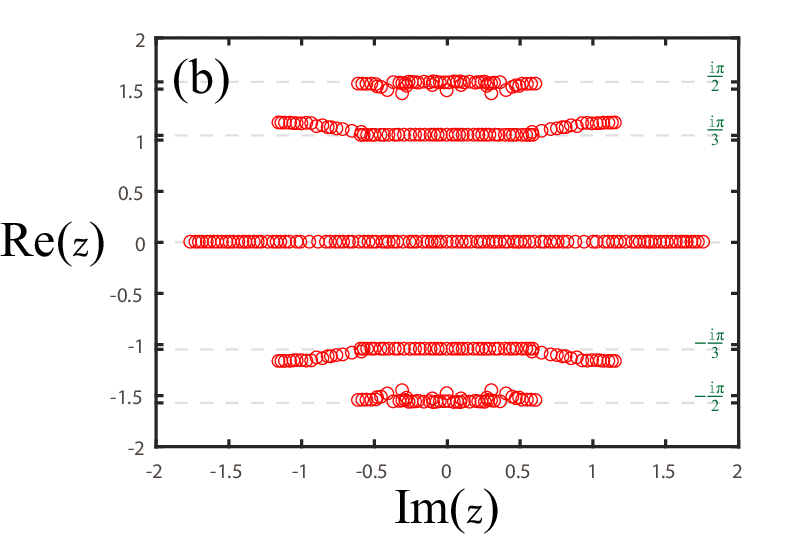}
    }
  \caption{Exact numerical results of $z$-roots (a) and $w$-roots (b) at any state with $N=8$ and $\eta=\frac{i\pi}{3}$.}\label{fig1}
\end{figure}
\noindent Moreover, the $t-W$ relation (\ref{t-W-relation-Eigen}) also implies that the $2N-1$ roots $\{z_j|j=1,\cdots,N-1\}$ and
$\{w_l|l=1,\cdots,N\}$ have to satisfy the zero points constraint equations
\bea
\hspace{-1.2truecm}&&\frac{\sinh^N(z_j+\frac{\eta}{2})\,\sinh^N(z_j-\frac{3}{2}\eta)}{\sinh^N\eta\,\sinh^N\eta} =\frac{\sinh^N(z_j-\frac{\eta}{2})}{\sinh^N\eta}\,W(z_j-\frac{\eta}{2}),\,\quad j=1,\cdots,N-1,\label{BAE-1}\\[6pt]
\hspace{-1.2truecm}&&\frac{\sinh^N(w_l+\eta)\,\sinh^N(w_l-\eta)}{\sinh^N\eta\,\sinh^N\eta}=-\Lambda(w_l)\,\Lambda(w_l-\eta),\, \quad l=1,\cdots,N.\label{BAE-2}
\eea

The corresponding eigenvalues of the Hamiltonian given by (\ref{xxzh})-(\ref{BC})  can be expressed  in terms of the $z$-roots $\{z_j|j=1,\cdots,N\}$ which are related to the zero points of $\Lambda(u)$  as
\bea
E=-2\sinh\eta\sum_{j=1}^{N-1}\coth(z_j-\frac{\eta}{2}) -N\cosh\eta.\label{Energy}
\eea
In the $\eta=\frac{i\pi}{3}$ case, we calculate the zero points of the eigenvalues $\Lambda(u)$ and $W(u)$ for any states by solving the constraint equations (\ref{Lam0}-\ref{BAE-2}). The results are shown in Fig.\ref{fig1}. From Fig.\ref{fig1} we observe the following $z$-roots and $w$-roots patterns
\begin{table}[!ht]
\centering
\caption{Patterns of $z$-roots and $w$-roots with $\eta=\frac{i\pi}{3}$.}
\begin{tabular}{|c|c|c|}
\hline &  {\bf $z$-roots patterns} & {\bf $w$-roots patterns}\\
\hline $\mathrm{I}$ & real $z_j^{(1)}$& real $w_l^{(1)}$ \\
\hline $\mathrm{II}$ & $-\frac{i\pi}{2}$ axis $z_j^{(2)}-\frac{i\pi}{2}$ with real $z_j^{(2)}$ & $\pm \frac{i\pi}{2}$ axis $w_l^{(2)}\pm\frac{i\pi}{2}$ with real $w_j^{(2)}$ \\
\hline $\mathrm{III}$ & conjugate pair $z_j^{(3)}\pm \eta$ with real $z_j^{(3)}$ & conjugate pair $w_l^{(3)}\pm \eta$ with real $w_j^{(3)}$\\
\hline
\end{tabular}
\label{zeros-table}
\end{table}

\begin{figure}[htbp]
  \centering
  \subfigure{\label{fig2:subfig:a} 
    \includegraphics[scale=0.367]{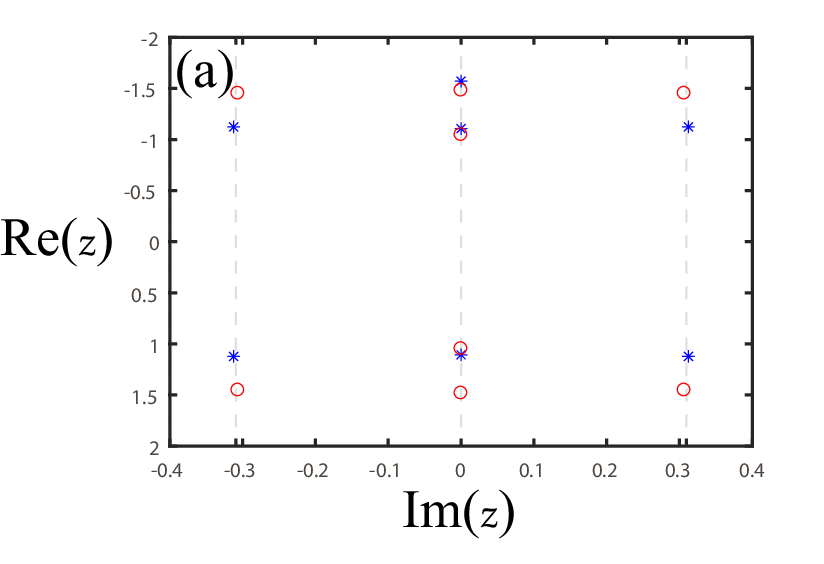}
    }
  \subfigure{\label{fig2:subfig:b} 
    \includegraphics[scale=0.367]{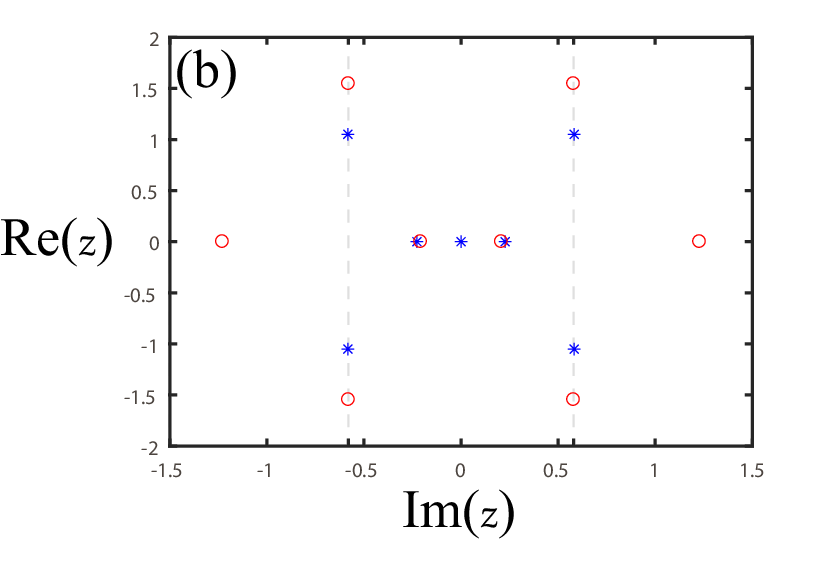}
    }
  \subfigure{\label{fig2:subfig:c} 
    \includegraphics[scale=0.367]{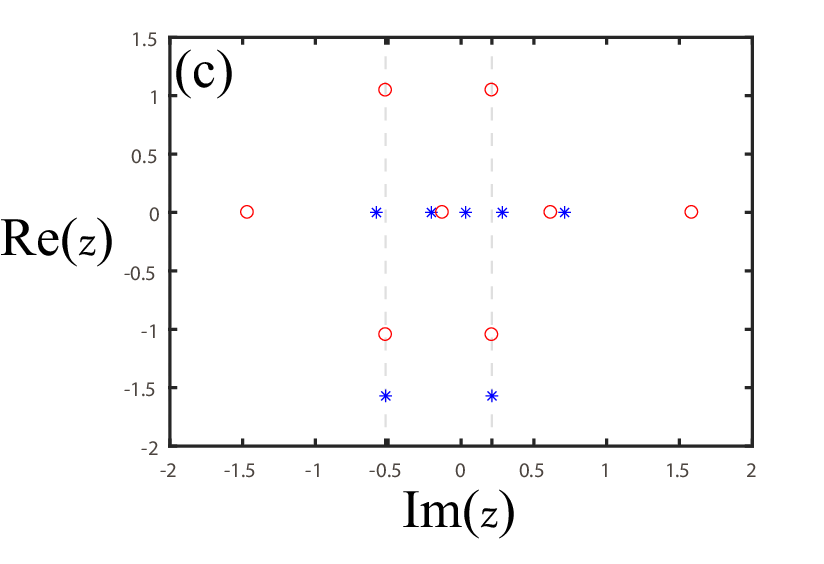}
    }
  \caption{Comparison of $z$-roots and $w$-roots at the same state with $N=8$ and $\eta=\frac{i\pi}{3}$. (a) the ground state; (b) the 10th excited state; (c) the 46th excited state. The blue
   asterisks indicate $z$-roots and the red circles specify $w$-roots.}\label{fig2}
\end{figure}

It can be shown that in the thermodynamic limit the following relationships between the $w$-roots and the $z$-roots hold (the proof is relegated to Appendix B)
\bea
\{w_l^{(2)}\}=\{z_j^{(3)}\},\qquad \{w_l^{(3)}\}=\{z_j^{(2)}\}.\label{zw-relation}
\eea
These relationships can also be verified from the exact numerical results of $z$-roots and $w$-roots at finite sites. For example, the roots of the ground state in Fig.\ref{fig2:subfig:a} show both relationships, while the roots of the 10th excited state in Fig.\ref{fig2:subfig:b} and the roots of the 46th excited state in Fig.\ref{fig2:subfig:c} demonstrate the 1st and 2nd relationships in (\ref{zw-relation}), respectively.


\section{Thermodynamics at the point of $\eta=\frac{i\pi}{3}$}
\label{Thermodyamics} \setcounter{equation}{0}
Based on the constraint equations (\ref{BAE-1}-\ref{BAE-2}) and the corresponding patterns of the $z$-roots and $w$-roots in the previous section, we in this section study the thermodynamics of the antiperiodic XXZ spin chain at the point of $\eta=\frac{i\pi}{3}$.

\subsection{Integral relations}
Denote $\bar{M}_1$ as the number of type I $w$-roots, and $M_1$, $M_2$ and $M_3$ as the numbers of types I, II, and III $z$-roots, respectively. The relations among $\bar{M}_1$, $M_1$, $M_2$ and $M_3$ are
\bea
&&\bar{M}_1+2M_2+2M_3=N,\\
&&M_1+2M_2+M_3=N-1.
\eea

Setting $z_j=z_j^{(1)}$ and $z_j=z_j^{(3)}+ \eta$ in constraint equations (\ref{BAE-1}), we obtain
\bea
\hspace{-1.3truecm}&&\frac{\sinh^N(z_j^{(1)}\hspace{-0.09truecm}+\hspace{-0.09truecm}\frac{\eta}{2})\,\sinh^N(z_j^{(1)}\hspace{-0.09truecm}-\hspace{-0.09truecm}\frac{3\eta}{2})}{ \sinh^N(z_j^{(1)}\hspace{-0.09truecm}-\hspace{-0.09truecm}\frac{\eta}{2})     } =W_{0}\prod_{n=1}^{\bar{M}_1}\sinh(z_j^{(1)}\hspace{-0.09truecm}-\hspace{-0.09truecm}w_n^{(1)}\hspace{-0.09truecm}-\hspace{-0.09truecm}\frac{\eta}{2} )\no\\[6pt]
\hspace{-1.3truecm}&&\times\prod_{k=1}^{M_3} \sinh(z_j^{(1)}\hspace{-0.09truecm}-\hspace{-0.09truecm}z_k^{(3)}\hspace{-0.09truecm}+\hspace{-0.09truecm}\eta )\sinh(z_j^{(1)}\hspace{-0.09truecm}-\hspace{-0.09truecm}z_k^{(3)}\hspace{-0.09truecm}+\hspace{-0.09truecm}\eta)\prod_{l=1}^{M_2} \sinh(z_j^{(1)}\hspace{-0.09truecm}-\hspace{-0.09truecm}z_l^{(2)}\hspace{-0.09truecm}+\hspace{-0.09truecm}\frac{\eta}{2})\sinh(z_j^{(1)}\hspace{-0.09truecm}-\hspace{-0.09truecm}z_l^{(2)}\hspace{-0.09truecm}-\hspace{-0.09truecm}\frac{3\eta}{2}),\label{BAEs-z1}
\eea
\bea
\hspace{-1.3truecm}&&\frac{\sinh^N(z_j^{(3)}\hspace{-0.09truecm}+\hspace{-0.09truecm}\frac{3\eta}{2})\,\sinh^N(z_j^{(3)}\hspace{-0.09truecm}-\hspace{-0.09truecm}\frac{\eta}{2})}{ \sinh^N(z_j^{(3)}\hspace{-0.09truecm}+\hspace{-0.09truecm}\frac{\eta}{2})     } =\,W_{0}\prod_{n=1}^{\bar{M}_1}\sinh(z_j^{(3)}\hspace{-0.09truecm}-\hspace{-0.09truecm}w_n^{(1)}\hspace{-0.09truecm}+\hspace{-0.09truecm}\frac{\eta}{2} )\no\\
\hspace{-1.3truecm}&&\times\prod_{k=1}^{M_3} \sinh(z_j^{(3)}\hspace{-0.09truecm}-\hspace{-0.09truecm}z_k^{(3)}\hspace{-0.09truecm}-\hspace{-0.09truecm}\eta )\sinh(z_j^{(3)}\hspace{-0.09truecm}-\hspace{-0.09truecm}z_k^{(3)}\hspace{-0.09truecm}-\hspace{-0.09truecm}\eta)\prod_{l=1}^{M_2} \sinh(z_j^{(3)}\hspace{-0.09truecm}-\hspace{-0.09truecm}z_l^{(2)}\hspace{-0.09truecm}+\hspace{-0.09truecm}\frac{3\eta}{2})\sinh(z_j^{(3)}\hspace{-0.09truecm}-\hspace{-0.09truecm}z_l^{(2)}\hspace{-0.09truecm}-\hspace{-0.09truecm}\frac{\eta}{2}),\label{BAEs-z3}
\eea
respectively. In the above derivation, we have used the property $\sinh(u+i\pi)=-\sinh(u)$ and the relation (\ref{zw-relation}). Dividing the complex conjugate of constraint equations (\ref{BAEs-z1}) (resp. constraint equations (\ref{BAEs-z3})) by (\ref{BAEs-z1}) (resp. (\ref{BAEs-z3})) and taking the logarithm of the resulting equations, we have
\bea
2\theta_1(z_j^{(1)})&=&
\frac{4\pi I_j^{(1)}}{N}-\frac{1}{N}\big[\sum_{n=1}^{\bar{M}_1}\theta_1(z_j^{(1)}-w_n^{(1)})\no\\[6pt]
& & +\sum_{k=1}^{M_3}2\theta_2(z_j^{(1)}-z_k^{(3)})+\sum_{l=1}^{M_2}\theta_1(z_j^{(1)}-z_l^{(2)})\big],\label{log-z1}\\
2\theta_1(z_j^{(3)})&=&
\frac{4\pi I_j^{(3)}}{N}-\frac{1}{N}\big[\sum_{n=1}^{\bar{M}_1}\theta_1(z_j^{(3)}-w_n^{(1)})\no\\[6pt]
& &+\sum_{k=1}^{M_3}2\theta_2(z_j^{(3)}-z_k^{(3)})+\sum_{l=1}^{M_2}\theta_1(z_j^{(3)}-z_l^{(2)})\big],\label{log-z3}
\eea
where $I_j^{(1)}$, $I_j^{(3)}$ denote the quantum numbers associated with the roots $z_j^{(1)}$, $z_j^{(3)}$, respectively, and $\theta_n(x)=2\cot^{-1}(\coth x \tanh\frac{n\gamma}{2})$.
Furthermore, multiplying the complex conjugate of constraint equations (\ref{BAEs-z1}) (resp. constraint equations (\ref{BAEs-z3})) by (\ref{BAEs-z1}) (resp. (\ref{BAEs-z3})) and taking the logarithm of the resulting equations, we obtain
\bea
\ln\hspace{-0.6truecm}&&|\sinh(z_j^{(1)}\hspace{-0.035truecm}-\hspace{-0.035truecm}\frac{3\eta}{2})|\hspace{-0.035truecm}=\hspace{-0.035truecm}\ln\,|W_{0}|\hspace{-0.035truecm}+\hspace{-0.035truecm}\frac{1}{N}\left[\sum_{n=1}^{\bar{M}_1}\ln|\sinh(z_j^{(1)}\hspace{-0.035truecm}-\hspace{-0.035truecm}w_n^{(1)}\hspace{-0.035truecm}-\hspace{-0.035truecm}\frac{\eta}{2})|\right.\no\\[6pt]
\hspace{-0.8truecm}&&\left.+\hspace{-0.035truecm}\sum_{k=1}^{M_3}2\ln|(z_j^{(1)}\hspace{-0.035truecm}-\hspace{-0.035truecm}z_k^{(3)}\hspace{-0.035truecm}+\hspace{-0.035truecm}\eta)|\hspace{-0.035truecm}+\hspace{-0.035truecm}\sum_{l=1}^{M_2}\ln|\sinh(z_j^{(1)}\hspace{-0.035truecm}-\hspace{-0.035truecm}z_l^{(2)}\hspace{-0.035truecm}+\hspace{-0.035truecm}\frac{\eta}{2})\sinh(z_j^{(1)}\hspace{-0.035truecm}-\hspace{-0.035truecm}z_l^{(2)}\hspace{-0.035truecm}-\hspace{-0.035truecm}\frac{3\eta}{2})|\right],\label{log-z12}\\
\ln\hspace{-0.6truecm}&&|\sinh(z_j^{(3)}\hspace{-0.035truecm}+\hspace{-0.035truecm}\frac{3\eta}{2})|\hspace{-0.035truecm}=\hspace{-0.035truecm}\ln\,|W_{0}|\hspace{-0.035truecm}+\hspace{-0.035truecm}\frac{1}{N}\left[\sum_{n=1}^{\bar{M}_1}\ln|\sinh(z_j^{(3)}\hspace{-0.035truecm}-\hspace{-0.035truecm}w_n^{(1)}\hspace{-0.035truecm}+\hspace{-0.035truecm}\frac{\eta}{2})|\right.\no\\[6pt]
\hspace{-0.8truecm}&&\left.+\hspace{-0.035truecm}\sum_{k=1}^{M_3}2\ln|(z_j^{(3)}\hspace{-0.035truecm}-\hspace{-0.035truecm}z_k^{(3)}\hspace{-0.035truecm}-\hspace{-0.035truecm}\eta)|\hspace{-0.035truecm}+\hspace{-0.035truecm}\sum_{l=1}^{M_2}\ln|\sinh(z_j^{(3)}\hspace{-0.035truecm}-\hspace{-0.035truecm}z_l^{(2)}\hspace{-0.035truecm}-\hspace{-0.035truecm}\frac{\eta}{2})\sinh(z_j^{(3)}\hspace{-0.035truecm}-\hspace{-0.035truecm}z_l^{(2)}\hspace{-0.035truecm}+\hspace{-0.035truecm}\frac{3\eta}{2})|\right].\label{log-z13}
\eea
In the thermodynamic limit $N\rightarrow\infty$, we define the densities of $z^{(l)}$-roots, $z^{(l)}$-holes, $w^{(l)}$-roots and $w^{(l)}$-holes per unit site as $\rho_l(z)$, $\rho_l^h(z)$, $\sigma_l(w)$ and $\sigma_l^h(w)$, respectively. Taking the continuum limits and derivatives of Eqs.(\ref{log-z1})-(\ref{log-z13}), we have
\bea
2a_1(z)\hspace{-0.2truecm}&=&\hspace{-0.2truecm}2\rho_1(z)+2\rho_1^{h}(z)-a_1*\sigma_1(z)+2a_2*\rho_3(z)+a_1*\rho_2(z),\label{de-z1}\\
2a_1(z)\hspace{-0.2truecm}&=&\hspace{-0.2truecm}2\rho_3(z)+2\rho_3^{h}(z)-a_1*\sigma_1(z)+2a_2*\rho_3(z)+a_1*\rho_2(z),\label{de-z3}\\
b_3(z)\hspace{-0.2truecm}&=&\hspace{-0.2truecm}b_1*\sigma_1(z)+2b_2*\rho_3(z)+(b_1+b_3)*\rho_2(z),\label{de-z13}
\eea
where $a_n(z)=\sin(n\gamma)/[\pi(\cosh 2z-\cos n\gamma)]$, $b_n(z)=\sinh(2z)/[\pi(\cosh 2z-\cos n\gamma)]$ and the symbol $*$ indicates convolution.

For the $z$-roots pattern (2) $z_j=z_j^{(2)}-\frac{i\pi}{2}$, the constraint equations (\ref{BAE-1}) can be rewritten as
\bea
\hspace{-1.5truecm}
&&\frac{\sinh^N(z_j^{(2)}\hspace{-0.09truecm}-\hspace{-0.09truecm}\eta)\,\sinh^N(z_j^{(2)})}{ \sinh^N(z_j^{(2)}\hspace{-0.09truecm}+\hspace{-0.09truecm}\eta)     } =\,W_{0}\prod_{n=1}^{\bar{M}_1}\sinh(z_j^{(2)}\hspace{-0.09truecm}-\hspace{-0.09truecm}w_n^{(1)}\hspace{-0.09truecm}+\hspace{-0.09truecm}\eta )\no\\[6pt]
\hspace{-1.5truecm}
&&\hspace{0.20truecm}\times\prod_{k=1}^{M_3} \sinh(z_j^{(2)}\hspace{-0.09truecm}-\hspace{-0.09truecm}z_k^{(3)}\hspace{-0.09truecm}-\hspace{-0.09truecm}\frac{\eta}{2} )\sinh(z_j^{(2)}\hspace{-0.09truecm}-\hspace{-0.09truecm}z_k^{(3)}\hspace{-0.09truecm}-\hspace{-0.09truecm}\frac{\eta}{2})\prod_{l=1}^{M_2} \sinh(z_j^{(2)}\hspace{-0.09truecm}-\hspace{-0.09truecm}z_l^{(2)}\hspace{-0.09truecm}-\hspace{-0.09truecm}\eta)\sinh(z_j^{(2)}\hspace{-0.09truecm}-\hspace{-0.09truecm}z_l^{(2)}).\label{BAEs-z2}
\eea
Due to the existence of the infinitesimal term $\prod_l\sinh(z_j^{(2)}-z_l^{(2)})$, we can only divide the complex conjugate of constraint equations (\ref{BAEs-z2}) by (\ref{BAEs-z2}) to eliminate the infinitesimal term. Taking the logarithm of the resulting equations, we obtain
\bea
2\theta_2(z_j^{(2)})=\hspace{-0.6truecm}&&\frac{4\pi I_j^{(2)}}{N}-\frac{1}{N}\big[\sum_{n=1}^{\bar{M}_1}\theta_2(z_j^{(2)}-w_n^{(1)})\no\\[6pt]
&&+\sum_{k=1}^{M_3}2\theta_1(z_j^{(2)}-z_k^{(3)})+\sum_{l=1}^{M_2}\theta_2(z_j^{(2)}-z_l^{(2)})\big],\label{log-z2}
\eea
where $I_j^{(2)}$ denote the quantum numbers associated with the roots $z_j^{(2)}$. Similarly, taking the derivative of Eq.(\ref{log-z2}) we have
\bea
2a_2(z)\hspace{-0.2truecm}&=&\hspace{-0.2truecm}\rho_2(z)+\rho_2^{h}(z)-a_2*\sigma_1(z)+2a_1*\rho_3(z)+a_2*\rho_2(z).\label{de-z2}
\eea

To determine the thermodynamics, we also need to obtain the density $\sigma_1(w)$.   Setting $w_j=w_j^{(1)}$ in constraint equations (\ref{BAE-2}), we obtain
\bea
&&\hspace{-1.2truecm}\sinh^N(w_j^{(1)}\hspace{-0.11truecm}+\hspace{-0.11truecm}\eta)\sinh^N(w_j^{(1)}\hspace{-0.11truecm}-\hspace{-0.11truecm}\eta)\hspace{-0.11truecm}=\hspace{-0.11truecm}\Lambda^{2}_{0}\prod_{n=1}^{M_1}\sinh(w_j^{(1)}\hspace{-0.11truecm}-\hspace{-0.11truecm}z_n^{(1)}\hspace{-0.11truecm}+\hspace{-0.11truecm}\frac{\eta}{2})\sinh(w_j^{(1)}\hspace{-0.11truecm}-\hspace{-0.11truecm}z_n^{(1)}\hspace{-0.11truecm}-\hspace{-0.11truecm}\frac{\eta}{2})\no\\
&&\hspace{0.10truecm}\times\prod_{k=1}^{M_3}\sinh(w_j^{(1)}\hspace{-0.11truecm}-\hspace{-0.11truecm}z_k^{(3)}\hspace{-0.11truecm}+\hspace{-0.11truecm}\frac{3\eta}{2})\sinh(w_j^{(1)}\hspace{-0.11truecm}-\hspace{-0.11truecm}z_k^{(3)}\hspace{-0.11truecm}+\hspace{-0.11truecm}\frac{\eta}{2})\sinh(w_j^{(1)}\hspace{-0.11truecm}-\hspace{-0.11truecm}z_k^{(3)}\hspace{-0.11truecm}-\hspace{-0.11truecm}\frac{\eta}{2})\sinh(w_j^{(1)}\hspace{-0.11truecm}-\hspace{-0.11truecm}z_k^{(3)}\hspace{-0.11truecm}+\hspace{-0.11truecm}\frac{3\eta}{2})\no\\
&&\hspace{0.10truecm}\times\prod_{l=1}^{M_2}\sinh(w_j^{(1)}\hspace{-0.11truecm}-\hspace{-0.11truecm}z_l^{(2)}\hspace{-0.11truecm}-\hspace{-0.11truecm}\eta)\sinh(w_j^{(1)}\hspace{-0.11truecm}-\hspace{-0.11truecm}z_l^{(2)}\hspace{-0.11truecm}+\hspace{-0.11truecm}\eta).
\eea
Taking the logarithm and derivative of this equation, we have
\bea
b_2(z)\hspace{-0.2truecm}&=&\hspace{-0.2truecm}b_1*\rho_1(z)+(b_1+b_3)*\rho_3(z)+b_2*\rho_2(z).\label{de-w1}
\eea
In addition, the expressions (\ref{Expansion-3})-(\ref{Expansion-4}) show that the total number of $z$-roots and $w$-roots must be $N-1$ and $N$ respectively, i.e.,
\bea
&&N\int_{-\infty}^{+\infty}[\rho_{1}(z)+\rho_{2}(z)+2\rho_{3}(z)]dz=N-1,\label{total-N-1}\\
&&N\int_{-\infty}^{+\infty}[\sigma_{1}(z)+2\rho_{2}(z)+2\rho_{3}(z)]dz=N.\label{total-N}
\eea
Now combining (\ref{de-z1})-(\ref{de-z13}) and (\ref{de-w1})-(\ref{total-N}), we obtain
\bea
\hspace{-1.2truecm}&&\sigma_1(z)=2g*[\rho_1(z)+\frac{1}{2N}(\delta(z-\infty)+\delta(z+\infty))]-\rho_2(z),\label{relation1}\\
\hspace{-1.2truecm}&&\rho_3^h(z) =\rho_1(z)+\frac{1}{2N}[\delta(z-\infty)+\delta(z+\infty)],\label{relation2}\\
\hspace{-1.2truecm}&&\rho_1^h(z) = \rho_3(z)+\frac{1}{2N}[\delta(z-\infty)+\delta(z+\infty)],\label{relation3}
\eea
where $g(z)=1/(2\gamma\cosh(3z))$ and $\delta(u)$ is the Dirac delta function. The existence of $\frac{1}{2N}(\delta(z-\infty)+\delta(z+\infty))$ is in order to satisfy the constraints (\ref{total-N-1}) and (\ref{total-N}). Using the relations (\ref{relation1})-(\ref{relation3}), after some calculations, we can rewrite (\ref{de-z1}) and (\ref{de-z2}) as
\bea
\rho_1(z)+\rho_3(z)\hspace{-0.2truecm}&=&\hspace{-0.2truecm}a_1(z)+a_2*\rho_1(z)-a_2*\rho_3(z)-a_1*\rho_2(z)\no\\[2pt]
\hspace{-0.2truecm}& &\hspace{-0.2truecm}+a_2*\frac{1}{2N}(\delta(z-\infty)+\delta(z+\infty))-\frac{1}{2N}(\delta(z-\infty)+\delta(z+\infty)),\\[2pt]
\rho_2(z)+\rho_2^h(z)\hspace{-0.2truecm}&=&\hspace{-0.2truecm}2g*[\rho_1(z)+\frac{1}{2N}(\delta(z-\infty)+\delta(z+\infty))].
\eea
Applying the Fourier transform and the inverse Fourier transform, we can express the densities $\rho_3(z)$ and $\rho_2^h(z)$ in terms of the densities $\rho_1(z)$ and $\rho_2(z)$ as follows
\bea
\rho_3(z)\hspace{-0.2truecm}&=&\hspace{-0.2truecm}g(z)+2g*g*[\rho_1(z)+\frac{1}{2N}(\delta(z-\infty)+\delta(z+\infty))]-\rho_1(z)\\
\hspace{-0.2truecm}& &\hspace{-0.2truecm}-\frac{1}{2N}(\delta(z-\infty)+\delta(z+\infty))-g*\rho_2(z)\label{p3}\\
\rho_2^h(z)\hspace{-0.2truecm}&=&\hspace{-0.2truecm}2g*[\rho_1(z)+\frac{1}{2N}(\delta(z-\infty)+\delta(z+\infty))]-\rho_2(z).\label{p2h}
\eea

\subsection{NLIEs and free eneegy}
\begin{figure}[htbp]
\centering
\includegraphics[scale=0.83]{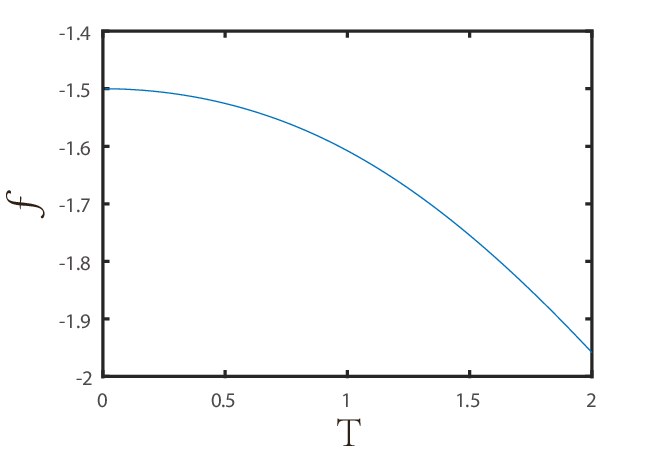}
  \caption{Free energy ($f$) vs Temperature for the anti-periodic XXZ chain.}\label{figfree}
\end{figure}

Thanks to the relations (\ref{relation2}) and (\ref{relation3}) of the densities $\rho_1(z)$ and $\rho_3(z)$, the number of the possible physical states in an infinitely small interval $[z,z+dz]$ of the $z$ space is given by
\bea
d\Omega(z)=\sum_{l=2}^3\frac{[N(\rho_{l}(z)+\rho_{l}^h(z))dz]!}{[N(\rho_{l}(z))dz]![N(\rho_{l}^h(z))dz]!}.
\eea
With the help of Stirling's formula $\ln N!\approx N\ln N-N$, we obtain the entropy in the interval
\bea
\hspace{-0.8truecm}dS(z)&&\hspace{-0.6truecm}=\ln d\Omega(z)\no\\
&&\hspace{-0.6truecm}\approx \sum_{l=2}^3 N{\{[\rho_{l}(z)+\rho_{l}^h(z)]\ln[\rho_{l}(z)+\rho_{l}^h(z)]}-\rho_{l}(z)\ln\rho_{l}(z)-\rho_{l}^h(z)\ln\rho_{l}^h(z)]\}dz.\label{entropy}
\eea
We define the relative density of the free energy as
\bea
f=\frac{F}{N}=\frac{E-TS}{N}=e-Ts.\label{free}
\eea
Using the energy expression (\ref{Energy}) and combining the $z$-roots patterns, the energy density is given by
\bea
e=\sqrt{3}\pi\int_{-\infty}^{+\infty}[a_1(z)\rho_{1}(z)-a_2(z)\rho_{2}(z)-a_1(z)\rho_{3}(z)]dz-\cosh(\eta).\label{energydensity}
\eea
Substituting (\ref{entropy}) and (\ref{energydensity}) into (\ref{free}), we have
\bea\label{free-int}
\hspace{-0.8truecm}f&=&
\sqrt{3}\pi\int_{-\infty}^{+\infty}\hspace{-0.075truecm}[a_1(z)\rho_{1}(z)\hspace{-0.075truecm}-\hspace{-0.075truecm}a_2(z)\rho_{2}(z)\hspace{-0.075truecm}-\hspace{-0.075truecm}a_1(z)\rho_{3}(z)]dz\hspace{-0.075truecm}-\hspace{-0.075truecm}\cosh(\eta)\nonumber\\
\hspace{-0.8truecm}& &
-T\sum_{i=2}^3\int_{-\infty}^{+\infty}\hspace{-0.075truecm}{ \{[\rho_{i}(z)\hspace{-0.075truecm}+\hspace{-0.075truecm}\rho_{i}^h(z)]\ln[\rho_{i}(z)\hspace{-0.075truecm}+\hspace{-0.075truecm}\rho_{i}^h(z)]}\hspace{-0.075truecm}-\hspace{-0.075truecm}\rho_{i}(z)\ln\rho_{i}(z)\hspace{-0.075truecm}-\hspace{-0.075truecm}\rho_{i}^h(z)\ln\rho_{i}^h(z)]\}dz.
\eea
For a thermal equilibrium state, the free energy should be minimized with respect to variations of $\rho_1(z)$ and $\rho_2(z)$ respectively, namely,
\bea
\frac{\delta f}{\delta \rho_1(z)}=0,\,\,\,\qquad \frac{\delta f}{\delta \rho_2(z)}=0.\label{free-delta}
\eea
Substituting (\ref{p3}), (\ref{p2h}) and (\ref{free-int}) into (\ref{free-delta}), and using $a_2(z)=a_1*g(z)$, we obtain
\beq
2\sqrt{3}\pi g(z)-T \{2g*g*\ln(1+\eta_3(z))+\ln(\eta_3^{-1}(z)) +2g*\ln(1+\eta^{-1}_2(z)) \}=0,\label{eta1}
\eeq
\beq
\ln(\eta_2(z))-g*\ln(1+\eta_3(z))=0,\label{eta2}
\eeq
where
\bea
\eta_l(z)=\frac{\rho_{l}^h(z)}{\rho_{l}(z)},\qquad l=2,3.
\eea
Putting (\ref{eta2}) in (\ref{eta1}) and combining with (\ref{relation1}), we can reduce (\ref{eta1}) and (\ref{eta2}) to
\bea
&&\ln(\eta_3(z))=-\frac{2\sqrt{3}\pi g(z)}{T}+ 2g*\ln(1+\eta_2(z)),\label{eta3}\\[6pt]
&&\ln(\eta_2(z))=g*\ln(1+\eta_3(z)).\label{eta2-1}
\eea
Thus $\eta_2(z)$ and $\eta_3(z)$ can be determined by iteration. Substituting (\ref{eta3}) and (\ref{eta2-1}) into (\ref{free-int}), we obtain the free energy of antiperiodic XXZ spin chain described by the Hamiltonian (\ref{xxzh})-(\ref{BC}) with $\eta=\frac{i\pi}{3}$ as
\bea
f=e_g-T\int_{-\infty}^{+\infty} {g(z) \ln(1+\eta_3(z))}dz,\label{free-1}
\eea
where $e_g=-\sqrt{3} \pi \int_{-\infty}^{+\infty}a_1(z)g(z)dz-\cosh(\eta)$ is the ground state energy density, which is the same as that of the periodic chain\cite{YY1966}.
The free energy vs the temperature $T$ is shown in Fig.\ref{figfree}. Based on the obtained free energy, other thermodynamic quantities such as specific heat can be calculated directly.


\section{Conclusions}
\label{Con}
In this paper, we have studied the thermodynamics of the antiperiodic XXZ spin chain with the anisotropic parameter $\eta=\frac{i\pi}{3}$ at finite temperature. We obtain the $t-W$ relations of the transfer matrix and the corresponding eigenvalues. We parameterize the eigenvalues of the transfer matrix and the fused transfer matrix by their zero points instead of Bethe roots. By substituting the zero points into the $t-W$ relation, the homogeneous zero points constraint equations are obtained. The patterns of distribution of the zero points at $\eta=\frac{i\pi}{3}$ are determined by solving the constraint equations. Based on these results, we have defined the densities of different zero points patterns and reconstructed entropy. Finally, we obtain the NLIEs and free energy describing the thermodynamics of the antiperiodic XXZ spin chain with the anisotropy parameter $\eta=\frac{i\pi}{3}$. Our results indicate that the twisted boundary has not effect on the free energy at the point of $\eta=\frac{i\pi}{3}$ in the thermodynamic limit.

\section*{Acknowledgments}

We thank Professor Yupeng Wang for valuable discussions. We acknowledge the financial support from the National Key R$\&$D Program of China (Grant No.2021YFA1402104), Australian Research Council Discovery Project DP190101529 and Future Fellowship FT180100099,
China Postdoctoral Science Foundation Fellowship 2020M680724, National Natural Science Foundation of China (Grant Nos. 12074410, 12247103, 12247179, 11934015 and 11975183), the Major Basic Research Program of Natural Science of Shaanxi Province (Grant Nos. 2021JCW-19 and 2017ZDJC-32), and the Strategic Priority Research Program of the Chinese Academy of Sciences (Grant No. XDB33000000).


\appendix
\section{Proof of the fusion relation}
\setcounter{equation}{0}
\renewcommand{\theequation}{A.\arabic{equation}}
In this appendix we prove the operator relation (\ref{t-W-relation-op}) between the transfer matrices by using the fusion technique \cite{Kul81, Kir86}.

For this purpose, let us first define $\{|i\rangle |i=1,2\}$ as an orthnormal basis of $\mathbb{C}^2$.
It is easy to know that the symmetric projection operator $P^{(+)}$ is a $3$-dimensional subspace spanned by the orthnormal basis $\{|11\rangle,\,\frac{1}{\sqrt{2}}(|12\rangle+|21\rangle),\,|22\rangle\}$, while the antisymmetric projection operator $P^{(-)}$ is an $1$-dimensional subspace spanned by
$\{\frac{1}{\sqrt{2}}(|12\rangle-|21\rangle)\}$.
The QYBE (\ref{QYB}) and the fusion condition ($R_{12}(-\eta)=-2P_{12}^{(-)}$) allow us to derive the relation
\bea
R_{23}(u)R_{13}(u\hspace{-0.09truecm}-\hspace{-0.09truecm}\eta)P^{(-)}_{12}
=P^{(-)}_{12}R_{23}(u)R_{13}(u\hspace{-0.09truecm}-\hspace{-0.09truecm}\eta)P^{(-)}_{12}
=\frac{\sinh(u\hspace{-0.09truecm}+\hspace{-0.09truecm}\eta)\sinh(u\hspace{-0.09truecm}-\hspace{-0.09truecm}\eta)}
{\sinh\eta\sinh\eta}\times {\rm id}.\label{Q-determinant}
\eea
Direct calculation shows that
\bea
P^{(+)}_{12}\,R_{23}(u)\,R_{13}(u-\eta)\,P^{(+)}_{12}
=\frac{\sinh u}{\sinh\eta}\, R^{(1,\frac{1}{2})}_{\{12\}\,3}(u),\label{Fusion-R}
\eea
where the fused $R$-matrix $R^{(1,\frac{1}{2})}_{\{12\}\,3}(u)$, in the basis $\{|11\rangle,\,\frac{1}{\sqrt{2}}(|12\rangle+|21\rangle),\,|22\rangle\}$, is given by
\bea
R^{(1,\frac{1}{2})}_{\{12\}\,3}(u)=\lt(\begin{array}{cccccc}
\frac{\sinh(u+\eta)}{\sinh\eta}&&&&&\\[6pt]
&\frac{\sinh(u-\eta)}{\sinh\eta}&\sqrt{2}&&&\\[6pt]
&\sqrt{2}\cosh\eta&\frac{\sinh u}{\sinh\eta}&&&\\[6pt]
&&&\frac{\sinh u}{\sinh\eta}&\sqrt{2}\cosh\eta&\\[6pt]
&&&\sqrt{2}&\frac{\sinh(u-\eta)}{\sinh\eta}&\\[6pt]
&&&&&\frac{\sinh(u+\eta)}{\sinh\eta}
\end{array}\rt).\label{Fusion-R-1}
\eea
The relation (\ref{Q-determinant}) leads to
\bea
P^{(-)}_{12}\sigma^{x}_{1}\sigma^{x}_{2}T_2(u)\,T_1(u-\eta)P^{(-)}_{12}=-a(u)\,d(u-\eta)\times {\rm id}.
\eea
Moreover, the relations (\ref{Q-determinant}) and (\ref{Fusion-R}) allow us to express the second term of the equation (\ref{Proof})
in terms of the fused monodromy matrix as
\bea
P^{(+)}_{12}\sigma^{x}_{1}\sigma^{x}_{2}T_2(u)\,T_1(u-\eta)P^{(+)}_{12}=\prod_{l=1}^{N}\frac{\sinh(u-\theta_l)}{\sinh\eta} \sigma_{\{12\}}^x\,T^{(1,\frac{1}{2})}_{\{12\}}(u),\label{Fused-Mono}
\eea
where the fused Pauli matrix is defined as
\bea
\sigma_{\{12\}}^x=\lt(\begin{array}{ccc}
                    0 & 0 & 1 \\
                    0 & 1 & 0 \\
                    1 & 0 & 0 \\
                  \end{array}\rt),
\eea
and the fused monodromy matrix $T^{(1,\frac{1}{2})}_{\{12\}}(u)$ can be expressed in terms of the fused $R^{(1,\frac{1}{2})}_{\{12\}\,3}(u)$ given by (\ref{Fusion-R})
\bea
T^{(1,\frac{1}{2})}_{\{12\}}(u)=R^{(1,\frac{1}{2})}_{\{12\} \,N}(u-\theta_N)\cdots R^{(1,\frac{1}{2})}_{\{12\} \,1}(u-\theta_1).\label{Fused-Mono-1}
\eea
Then the fused transfer matrix $\mathbb{W}(u)$ is constructed as follows
\bea
\mathbb{W}(u)=tr_{\{12\}}\lt\{\sigma_{\{12\}}^x T^{(1,\frac{1}{2})}_{\{12\}}(u)\rt\},
\eea
where $tr_{\{12\}}$ denotes trace over the subspace $P^{(+)}$.
From the definitions of (\ref{Fusion-R-1}) and (\ref{Fused-Mono-1}), we know that the matrix elements of $\mathbb{W}(u)$ are $N$-degree operator-valued trigonometric polynomials of $u$. This completes the proof of the fusion relation (\ref{t-W-relation-op}).


\section{Proof of the relation between $z$-roots and $w$-roots }
\setcounter{equation}{0}
\renewcommand{\theequation}{B.\arabic{equation}}
In this appendix we prove the relations (\ref{zw-relation}) by using the $t-W$ relation (\ref{t-W-relation-Eigen}). Setting $u=z_j-\frac{\eta}{2}$ and $u=z_j+\frac{\eta}{2}$ in (\ref{t-W-relation-Eigen}), respectively, we obtain
\bea
\frac{\sinh^N(z_j+\frac{\eta}{2})\,\sinh^N(z_j-\frac{3\eta}{2})}{\sinh^N\eta\,\sinh^N\eta}  =\frac{\sinh^N(z_j-\frac{\eta}{2})}{\sinh^N\eta}\,W_{0}
\prod_{l=1}^N \frac{\sinh(z_j-w_l-\frac{\eta}{2})}{\sinh\eta},\label{t-W-B1}\\[6pt]
\frac{\sinh^N(z_j+\frac{3\eta}{2})\,\sinh^N(z_j-\frac{\eta}{2})}{\sinh^N\eta\,\sinh^N\eta}=\frac{\sinh^N((z_j+\frac{\eta}{2})}{\sinh^N\eta}\,W_{0}
\prod_{l=1}^N \frac{\sinh(z_j-w_l+\frac{\eta}{2})}{\sinh\eta}.\label{t-W-B2}
\eea
Dividing (\ref{t-W-B2}) by (\ref{t-W-B1}), we have
\bea
\frac{\sinh^{2N}(z_j-\frac{\eta}{2})}{\sinh^{2N}(z_j+\frac{\eta}{2})} =
\prod_{l=1}^N \frac{\sinh(z_j-w_l+\frac{\eta}{2})}{\sinh(z_j-w_l-\frac{\eta}{2})}.\label{t-W-B3}
\eea
For complex $z$-roots $z_j$ with negative imaginary part and ${\rm Im}(z_j)>- \frac{\pi}{2}$, we readily have
\bea
\Big|\sinh(z_j-\frac{\eta}{2})\Big|>\Big|\sinh(z_j+\frac{\eta}{2})\Big|.\label{t-W-B4}
\eea
This indicates that the module of the left hand side of (\ref{t-W-B3}) is bigger than 1. Thus in
the thermodynamic limit $N\rightarrow\infty$, the left hand side tends to infinity exponentially. To keep (\ref{t-W-B3}) holding,  the right hand side of (\ref{t-W-B3}) must also approach infinity in the same order.  Thus the denominator of the first term in the right hand side must tend to zero exponentially, which leads to $z_j-w_l-\frac{\eta}{2}\rightarrow 0$. Similarly, we can obtain $z_j-w_l+\frac{\eta}{2}\rightarrow 0$ for  complex $z$-roots $z_j$ with positive imaginary part. Therefore, for the $z$-roots patterns (3) conjugate pair $z_j^{(3)}\pm \eta$, the $w$-root structure satisfies $w_l=z_j^{(3)}\pm \frac{i\pi}{2}=w_j^{(2)}\pm \frac{i\pi}{2}$.

Next, we consider the $z$-roots patterns (2) $-\frac{i\pi}{2}$ axis $z_j^{(2)}-\frac{i\pi}{2}$. Putting $z_j=z_j^{(2)}-\frac{i\pi}{2}$ in (\ref{t-W-B1}), we have
\bea
\frac{\sinh^N(z_j-\eta)\,\sinh^N(z_j)}{\sinh^N(z_j+\eta)}  =\,W_{0}
\prod_{l=1}^N \sinh(z_j-w_l+\eta).\label{t-W-B5}
\eea
Multiplying the complex conjugate of constraint equations (\ref{t-W-B5}) by (\ref{t-W-B5}) and taking the logarithm of the resulting equation, we obtain
\bea
2N\ln|\sinh(z_j^{(2)})|=2\ln|W_0|+\prod_{l=1}^N \ln| \sinh(z_j^{(2)}-w_l+\eta)\sinh(z_j^{(2)}-w_l^{*}-\eta)|.\label{t-W-B6}
\eea
When we take the continuum limit of (\ref{t-W-B6}) in the thermodynamic limit, the left hand side of (\ref{t-W-B6}) will contribute a $\delta$-function.  To keep (\ref{t-W-B6}) holding, the antilogarithm of the second term in the right hand side must tend to zero exponentially, which leads to $z_j^{(2)}-w_l+\eta\rightarrow 0$ and $z_j^{(2)}-w_l^{*}-\eta\rightarrow 0$. Therefore, for the $z$-roots pattern (2) $\frac{i\pi}{2}$ axis $z_j^{(2)}-\frac{i\pi}{2}$, the $w$-root structure satisfies $w_j=z_j^{(2)}\pm\eta=w_j^{(3)}\pm \eta$. As a result,  the relation (\ref{zw-relation}) is achieved. This completes our proof.

\end{document}